\documentclass[a4paper,twocolumn,brazil,english,aps,noshowpacs]{revtex4-1}
\usepackage[T1]{fontenc}
\usepackage[utf8]{inputenc}
\usepackage{verbatim}
\usepackage{bm}
\usepackage{amsmath}
\usepackage{amssymb}
\usepackage{stmaryrd}
\usepackage{graphicx}
\usepackage{esint}
\usepackage{color}

\makeatletter

\pdfpageheight\paperheight
\pdfpagewidth\paperwidth

\everymath{\displaystyle}

\makeatother

\usepackage{babel}
\begin{document}

\title{ Fermi points and topological quantum phase transitions in a model of superconducting wires }

\author{T.O. Puel$^{1}$}

\email{tharnier@me.com}

\selectlanguage{english}%

\author{P.D. Sacramento$^{1,2}$}

\author{M. A. Continentino$^{1}$}

\affiliation{$^{1}$Centro Brasileiro de Pesquisas Físicas, Rua Dr. Xavier Sigaud,
150, Urca 22290-180, Rio de Janeiro, RJ, Brazil }

\affiliation{$^{2}$CeFEMA, Instituto Superior Técnico, Universidade de Lisboa,
Av. Rovisco Pais, 1049-001 Lisboa, Portugal}

\date{\today}
\begin{abstract}
The importance of models with an exact solution for the study of materials with non-trivial topological properties
has been extensively demonstrated. Among these, the Kitaev model of a one-dimensional p-wave superconductor
plays a guiding role in the search for Majorana modes in condensed matter systems.
Also, the  $sp$ chain, with an anti-symmetric mixing among the $s$ and $p$ bands provides a paradigmatic example of a topological insulator with well understood properties.
There is an intimate relation
between these two models and in particular their topological quantum phase transitions share the same universality class.
Here we consider a two-band $sp$ model of spinless fermions with an attractive   (inter-band) interaction. Both the interaction and hybridization between the $s$ and $p$ fermions are anti-symmetric.
The zero temperature phase diagram of
the model presents a variety of phases including a Weyl superconductor,  topological insulator and trivial phases.
The quantum phase transitions between these phases can be either continuous or discontinuous. We show that the transition from the topological superconducting phase to the trivial one has critical exponents different from those of an equivalent transition in Kitaev's model.
\end{abstract}


\maketitle

\section{Introduction}

Since the first strong experimental evidence of Majorana fermions~\cite{Mourik-Zuo-Plissard-Bakkers-Kouwenhoven-2012}
in a hybrid superconductor-semiconductor one dimensional system, the
search for exotic states supporting Majorana fermions has attracted
increasing interest in condensed matter physics. Recent observations
have reinforced the existence of Majorana, specially the one made
in ferromagnetic atomic chains on a superconductor (SC)~\cite{Perge-Drozdov-Li-Chen-Jeon-Seo-MacDonald-Bernevig-Yazdani-2014}.
Anomalous behaviour on this experimental evidence~\cite{Finck-Harlingen-Mohseni-Jung-Li-2013}
indicates that the appearance of Majorana may have yet unknown sources.
Also some technical difficulties such as highly localized states compared to the material parameters~\cite{Peng-Pientka-Glazman-Oppen-2015}
or high temperatures that prevent to have access inside the gap~\cite{Dumitrescu-Roberts-Tewari-Sau-Sarma-2015}
have left the existence of Majorana inconclusive.
The running for the experimental discovery of Majorana is well described
in~\cite{Elliott-Franz-2015}.

It is well known that the Kitaev model~\cite{Kitaev-2001,Kitaev-2003,Alicea-2012}
-- anti-symmetric pairs of spinless fermions in 1D -- exhibits a non-trivial
topological phase with Majorana modes at the ends of a $p$-wave superconducting
chain. The excitations at the ends of the chain depend on the quantum
state of the system, which in turn is determined by the ratio $\mu/2t$,
between the chemical potential $\mu$ and the hopping $t$. If $|\mu|/2t<1$
the chain is superconducting with non-trivial topological properties.
This \textit{weak pairing} phase presents Majorana fermions at its
ends. Otherwise, if $|\mu|/2t>1$ the chain is in a \textit{strong coupling}
superconducting phase with trivial topological properties and has
no end states~\cite{Alicea-2012}.

The importance of the mixing of $sp$ bands for topological insulators
has already been pointed out in different contexts, including that
of the spin quantum Hall effect~\cite{Bernevig-Hughes-Zhang-2006}
and a cold atom version of the $sp$-chain, the $sp$-ladder\cite{Li-Zhao-Liu-2013}.
The $sp$ mixing is in a special class that mixes orbitals with angular momenta that differ by an odd number.
This implies the anti-symmetric property $V(-\mathbf{r})=-V(\mathbf{r})$ or in momentum space
$V(-\mathbf{k})=-V(\mathbf{k})$~\cite{Continentino-Deus-Padilha-Caldas-2014}.

In addition, it was recently shown~\cite{Dzero-Sun-Galitski-Coleman-2010,Continentino-Caldas-Nozadze-Trivedi-2014,Alexandrov-Coleman-2014}
an intimate relation between a two band insulator with anti-symmetric
hybridization and the Kitaev model, as regards to the topological
properties and their end states. By tuning the parameters of the 1D
$sp$ chain, the system can be driven, through a topological quantum phase transition,
from a trivial to a topological insulator. As a result they found
two Majorana zero modes at the ends of an insulating chain. In the
search for Majorana modes, Kitaev's model is a most clear example of the importance of exactly soluble
models as guides in this difficult path.

Here we consider a different
model of  a $p$-wave superconducting chain that also can be solved exactly.
This study will throw further light on the role of topology and different
type of interactions and symmetry breaking terms as conditions for
the existence of Majorana in superconducting wires. We consider a
two-band model of spinless fermions in a chain with inter-band attractive
interactions and an anti-symmetric hybridization.  The
model is exactly soluble and we obtain its zero temperature phase diagram.
For a fixed small value of the mixing, as the ratio $\mu/2t$ of the model increases, such that $|\mu|/2t>1$,  there is a topological quantum phase transition
from a gapless (topologically non-trivial) to a gapped (trivial) superconducting phase. The phase diagram resembles that of Kitaev's model, however the nature of the topological phases, as well as, the topological transition are distinct.
In our model the non-trivial superconducting phase has
gapless Fermi points. These gapless points have the characteristic
of Weyl fermions in 3D systems~\cite{Liu-Hu-Pu-2015,Cao-Zou-Liu-Yi-Long-Hu-2014,Yang-Pan-Zhang-2014},
as they have a non-degenerated linear dispersion relation and appear or disappear
in pairs, only when two Fermi points unite. The conservation of the $\mathit{topological charge}$ associated with
these Fermi points~\cite{Volovik-2009} confers a non-trivial topological character for
this phase. Furthermore, we show that the topological quantum phase transition at $(|\mu|/2t)_c=1$
is in a different universality class of that of Kitaev's model.
Next, fixing the chemical potential, say at $\mu=0$, and increasing the hybridization the system is driven from
the Weyl superconducting phase (WSC) to a topological insulator through a first order quantum phase transition.
Topological phase transitions are known  to produce anomalies in thermodynamic quantities\cite{Seo-Zhang-Tewari-2013} and we obtain these here, with special emphasis on the behaviour of the compressibility.

In section \ref{sec:the model} we define the model and analyse the
superconductivity stability. In section \ref{sec:phase diagram}
we show the phase diagram for the superconducting stability.
We also discuss the nature of the transitions
and show the localization of end states in a finite system. In section
\ref{sec:Topological-condition} we calculate topological invariants of different phases,
and discuss the topological properties of the model in Majorana basis.
Finally, in section \ref{sec:Conclusions}
we present some conclusions and highlight the main results.

\section{Defining the model\label{sec:the model}}

We consider a two-band problem with hybridisation and triplet inter-band
superconductivity in 1D, i.e., a chain with two orbitals per site, with angular momenta differing by an odd number,
let's say $p$ and $s$. The pairing between fermions on different
bands (inter-band) is always $p$-wave kind, in the sense that the
pairing of spinless fermions is anti-symmetric.
The problem can be viewed as a generalization of Kitaev's model to
two orbitals and only interband pairing.
We also have the anti-symmetric hybridisation term that, under some
conditions, was shown to be responsible for topological phases\cite{Dzero-Sun-Galitski-Coleman-2010,Continentino-Caldas-Nozadze-Trivedi-2014,Alexandrov-Coleman-2014}.
The simplest Hamiltonian in the momentum space that describes those
types of superconductivity and hybridization can be written as
\begin{eqnarray}
\label{hamiltonian}
\mathcal{H} & = & \sum_{k}\left\{ -\mu\left(c_{k}^{\dagger}c_{k}+p_{k}^{\dagger}p_{k}\right)+2t\cos\left(k\right)\left(p_{k}^{\dagger}p_{k}-c_{k}^{\dagger}c_{k}\right)\right.\nonumber \\
 &  & \qquad\left.-i\Delta\sin\left(k\right)c_{k}^{\dagger}p_{-k}^{\dagger}+iV\sin(k)c_{k}^{\dagger}p_{k}+\text{h.c.}\right\} ,\nonumber \\
\end{eqnarray}
where $\mu$ is the chemical potential, $\Delta$ is the $sp$ pairing
amplitude, and $V$ is the anti-symmetric hybridization amplitude.
Note that the hopping amplitude $t$ has different sign in each band,
representing particles for the orbital $s$ and holes for the orbital
$p$. We can write the same Hamiltonian using the Bogoliubov-de Gennes
(BdG) representation as

\begin{equation}
\mathcal{H}=\sum_{k}\boldsymbol{C}_{k}^{\dagger}\mathcal{H}_{k}\boldsymbol{C}_{k},
\end{equation}
with $\boldsymbol{C}_{k}^{\dagger}=\left(c_{k}^{\dagger}p_{k}^{\dagger}c_{-k}p_{-k}\right)$
and
\begin{equation}
\mathcal{H}_{k}=-\mu\Gamma_{z0}-\varepsilon_{k}\Gamma_{zz}+\Delta_{k}\Gamma_{yx}-V_{k}\Gamma_{zy},\label{eq:hamilt k basis}
\end{equation}
where $\Gamma_{ab}=r_{a}\varotimes\tau_{b},\; \forall\;a,b=x,y,z$,
and $r_{x,y,z}/\tau_{x,y,z}$ are the Pauli matrices acting on particle-hole/orbitals space, respectively,
and $r_{0}=\tau_{0}$ are the $2\times2$ identity matrix. We have defined $\varepsilon_{k}=2t\cos\left(k\right)$,
$V_{k}=V\sin\left(k\right)$, and $\Delta_{k}=\Delta\sin\left(k\right)$.

\subsection{Energy spectrum\label{sub:Energy-spectrum}}

\begin{figure*}
\hfill{}\includegraphics[width=0.3\textwidth]{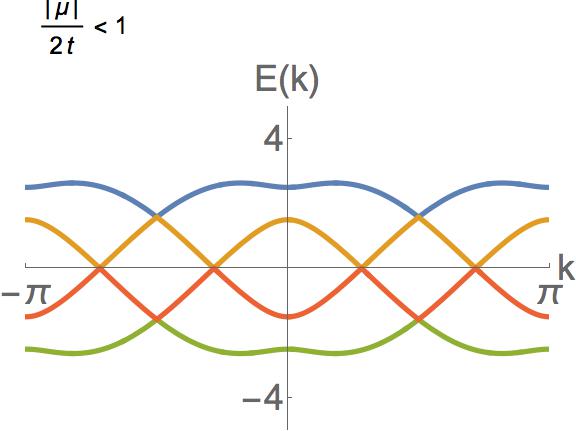}\hfill{}\includegraphics[width=0.3\textwidth]{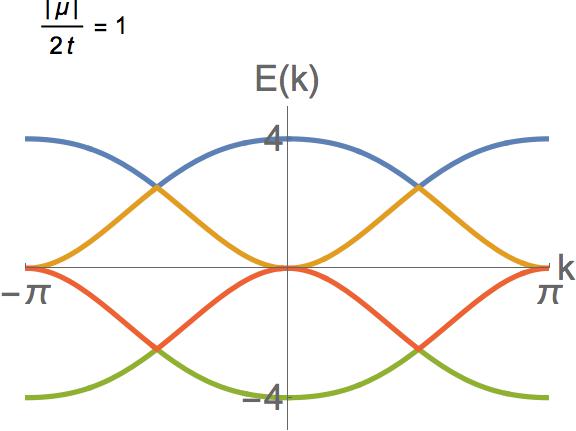}\hfill{}
\includegraphics[width=0.3\textwidth]{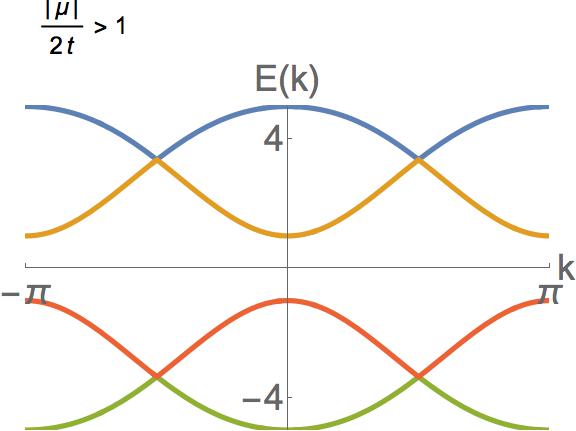}\hfill{}
\protect\caption{\label{fig:spectrum} Dispersion relations for a pure $sp$ superconducting
system, without hybridization, for $\Delta=1.5$ and $V=0$. The system
supports gapless excitations when $\left(\left|\mu\right|/2t\right)<1$
and is gapped otherwise, specifically, we set $\left(\left|\mu\right|/2t\right)=0.25$
(left) and $\left(\left|\mu\right|/2t\right)=1.5$ (right).
In this work we show that both regions
have self-consistent solutions for the superconductivity. The central
panel shows the zero modes annihilation; note that it happens in pairs
of nodes.}
\end{figure*}

Since a topological phase transition only occurs when a gap closes,
looking for gapless points on the energy spectrum may indicate this
transition. The model considered here has the following energy dispersion
relations,
\begin{equation}
E(k)=\pm\sqrt{Z_{1}\pm2\sqrt{Z_{2}}},\label{eq:dispersion relation}
\end{equation}
where $Z_{1}=A(k)+B\left(k\right)$ and $Z_{2}=A(k)B\left(k\right)$,
with $A(k)=\varepsilon_{k}^{2}+V_{k}^{2}$ and $B\left(k\right)=\Delta_{k}^{2}+\mu^{2}$.
Looking for gapless points ($E(k)=0$) the possible solution is $A(k)=B\left(k\right)$,
i.e.,
\begin{eqnarray}
\mu^{2} & = & \varepsilon_{k}^{2}+V_{k}^{2}-\Delta_{k}^{2}\nonumber \\
 & = & \left[\left(2t\right)^{2}\cos^{2}\left(k\right)+\left(V^{2}-\Delta^{2}\right)\sin^{2}\left(k\right)\right].\label{eq:condicao fechamento gap}
\end{eqnarray}
We will analyze the equation above more deeply in section \ref{sec:phase diagram}.
First we would like to highlight the case with no hybridization, $V=0$,
in which the system is always gapless whenever $\left|\mu\right|\leq2t$.
The existence of these gapless modes represents a substantial difference
between this and the Kitaev model, see figure \ref{fig:spectrum}.
We will see in the next section that even in this non-gapped region
the system shows superconductivity. On the other side, when $\left|\mu\right|>2t$,
the system is fully gapped but superconductivity is still present up to $\left|\mu\right|<4t$.

Deep inside the gapless phase the crossings between bands have a linear
dispersion relation (\ref{fig:spectrum}a) and define Dirac nodes.
Furthermore, we note that the bands are non-degenerate and the nodes
appear and disappear only when two nodes are combined, as one can see comparing
figures \ref{fig:spectrum}a and \ref{fig:spectrum}b. This is a characteristic
of Weyl fermions in 3D or 2D SC\cite{Yang-Pan-Zhang-2014} and in
topological superfluidity\cite{Liu-Hu-Pu-2015,Cao-Zou-Liu-Yi-Long-Hu-2014}.
In this sense, the model here presented can be called 1D Weyl SC.

\subsection{Self-consistent equations for the superconductivity and the occupation
number}

We may calculate the self-consistent inter-band superconducting order
parameter, $\Delta$, from the gap equation
\begin{eqnarray}
\Delta & = & -\frac{4g}{L}\sum_{k}i\sin\left(k\right)\left\langle p_{-k}c_{k}\right\rangle ,
\end{eqnarray}
where, $g$ is the attractive energy between the spinless fermions,
$L$ is the length of the chain, and the correlation function $\left\langle p_{-k}c_{k}\right\rangle $
is obtained from the fluctuation-dissipation theorem\cite{Tyablikov-1967}
(a similar calculation was recently done in \cite{Continentino-Deus-Padilha-Caldas-2014}),
such that
\begin{equation}
\left\langle p_{-k}c_{k}\right\rangle =\frac{i}{2\pi}\int f\left(\omega\right)\left[\left\langle \left\langle c_{k},p_{-k}\right\rangle \right\rangle ^{r}-\left\langle \left\langle c_{k},p_{-k}\right\rangle \right\rangle ^{a}\right]d\omega;
\end{equation}
the Green's functions $\left\langle \left\langle c_{k},p_{-k}\right\rangle \right\rangle ^{r,a}$
(retarded and advanced) are obtained from the Greenian operator \cite{Foo-Johnson-1976,Foo-Thorpe-Weaire-1976,Foo-Wong-1974},
i.e., $\left\langle \left\langle \boldsymbol{G}\right\rangle \right\rangle _{k}=\left(\omega\mathbb{I}_{4\times4}-\mathcal{H}_{k}\right)^{-1}$,
and $f\left(\omega\right)$ is the Fermi distribution. If we proceed
with the calculations, see appendix \ref{app:Gap-equation}, the gap
equation at T=0 becomes
\begin{eqnarray}
\frac{1}{g} & = & \frac{1}{L}\sum_{k}\frac{4\sin^{2}\left(k\right)}{\left(\omega_{1}+\omega_{2}\right)}\delta_{k},\label{eq:condition}
\end{eqnarray}
where
\begin{equation}
\delta_{k}=\begin{cases}
1 & \quad\text{if}\quad B\left(k\right)>A\left(k\right),\\
0 & \quad\text{otherwise}.
\end{cases}\label{eq:constrain}
\end{equation}
Also, $\omega_{1}$ and $\omega_{2}$ are the eigenvalues of the Hamiltonian,
such that,
\begin{equation}
\begin{array}{c}
\omega_{1}(k)\equiv E\left(k\right)_{+}=\sqrt{Z_{1}+2\sqrt{Z_{2}}},\\
\omega_{2}(k)\equiv E\left(k\right)_{-}=\sqrt{Z_{1}-2\sqrt{Z_{2}}}.
\end{array}
\end{equation}

We can verify the stability of the superconducting phase calculating
the parameters $\Delta$ and $\mu$ self-consistently from the gap
and  the occupation number equation given by,
\begin{equation}
n=n_{s}+n_{p}=\frac{1}{L}\sum_{k}\left[\left\langle c_{k}^{\dagger}c_{k}\right\rangle +\left\langle p_{k}^{\dagger}p_{k}\right\rangle \right],
\end{equation}
with $n_{s}$ and $n_{p}$ the occupation numbers for the $s$ and
$p$ bands, respectively. For the model considered here, the above
equation is
\begin{equation}
n=\frac{1}{L}\sum_{k}\left[\frac{\mu}{\left(\omega_{1}+\omega_{2}\right)}\delta_{k}+\frac{1}{2}\right],\label{eq:auto-consistentes-sem-hopping}
\end{equation}
where $0<n<2$ is the total occupation number per site of the chain.

\section{Phase Diagram\label{sec:phase diagram}}

The solution of the coupled self-consistent equations for the gap
and the chemical potential is complicated by the constraints of the
sums in momentum space (Eq. (\ref{eq:constrain})). In this section we
present the phase diagram of our model system obtained directly from
a numerical solution of the BdG equations fixing the chemical potential.
The Hamiltonian defined in Eq. (\ref{eq:hamilt k basis}) can be solved using BdG transformations
as
\begin{equation}
\begin{array}{c}
c_{k}=\sum_{n}\left[u_{n,k}^{s}\gamma_{n,k}+\left(v_{n,k}^{s}\right)^{*}\gamma_{n,-k}^{\dagger}\right],\\
p_{k}=\sum_{n}\left[u_{n,k}^{p}\gamma_{n,k}+\left(v_{n,k}^{p}\right)^{*}\gamma_{n,-k}^{\dagger}\right].
\end{array}
\end{equation}
This transformation diagonalizes the Hamiltonian in the form
\begin{equation}
\mathcal{H}_{k}\begin{pmatrix}u_{n,k}^{s}\\
u_{n,k}^{p}\\
v_{n,-k}^{s}\\
v_{n,-k}^{p}
\end{pmatrix}_{n}=E_{n}\begin{pmatrix}u_{n,k}^{s}\\
u_{n,k}^{p}\\
v_{n,-k}^{s}\\
v_{n,-k}^{p}
\end{pmatrix}_{n},
\end{equation}
where $E_{n}$ are the energy eigenvalues and the wave functions spinor
are the eigenstates. The self-consistent solution implies that the
pairing can be obtained using
\begin{eqnarray}
\Delta & = & -\frac{2g}{L}\sum_{k}i\sin\left(k\right)\left(\left\langle c_{-k}p_{k}\right\rangle +\left\langle p_{-k}c_{k}\right\rangle \right).
\end{eqnarray}
At zero temperature, using the representation of the fermionic operators
in terms of the wave functions and the Bogoliubov coefficients we
may write
\begin{eqnarray}
\Delta & = & \frac{2g}{L}\sum_{k}\sum_{n}i\sin\left(k\right)\left[u_{n,k}^{s}\left(v_{n,-k}^{p}\right)^{*}+u_{n,k}^{p}\left(v_{n,-k}^{s}\right)^{*}\right].\nonumber \\
\end{eqnarray}

\begin{figure*}
\includegraphics[width=0.3\textwidth]{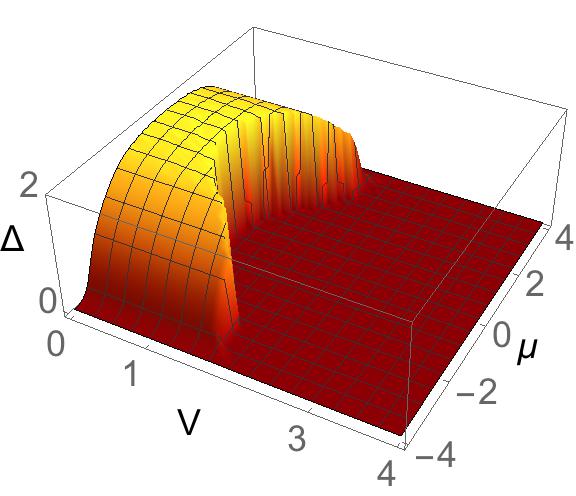}
\includegraphics[width=0.23\textwidth]{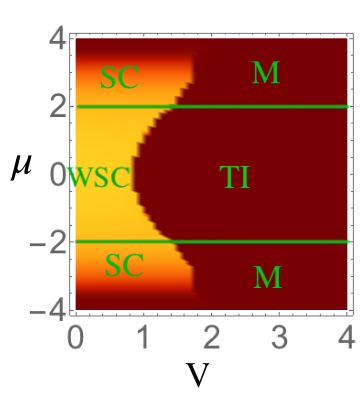}
\includegraphics[width=0.3\textwidth]{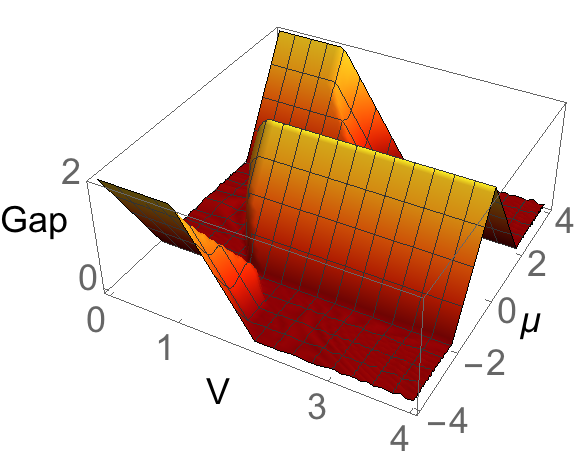}
\protect\caption{\label{fig1} The superconducting amplitude $\Delta$ (left panel)
and spectrum gap (right panel) for different values of $V$ and $\mu$,
with $g=1.7$. The middle panel
shows all the different phases, that are: trivial superconducting
gapped phase (SC), metallic (M), gapless Weyl superconductor
(WSC) and topological insulator (TI).}
\end{figure*}

In Fig. \ref{fig1}a we show the numerical results for the order parameter
$\Delta$ as a function of the chemical potential and hybridization
for a fixed value of the attractive interaction $g=1.7$. All quantities
are normalized by the hopping term $t$. In Fig.~\ref{fig1}c we show
the gap for excitations for the same range of parameters.
The results in these figures allow us to obtain the zero temperature
phase diagram of the system shown in Fig.~\ref{fig1}b.

In agreement with our previous discussions we find a gapless superconducting phase for
$\left(\left|\mu\right|/2t\right)<1$ and $V/2t<V_{c}(\mu)/2t \equiv [(\mu/2t)^{2} + (\Delta_0/2t)^{2}]^{1/2}$
named Weyl SC ($WSC$) in the phase diagram.
We note that at the transition ($V=V_c(\mu)$) the gapless points always occur at $k = \pm \pi /2$.
The quantity $\Delta_0=\Delta_0(\mu)$ is the value of the order parameter at $V=0$ for a given chemical potential value $\mu$.

For $\left(\left|\mu\right|/2t\right)>1$ and $V<V_{c}(\mu)$ the system presents a gapped
superconducting phase with trivial topological properties similar to the strong coupling superconducting phase of
Kitaev's model. In this phase, named $SC$ in the phase
diagram, the order
parameter vanishes continuously as the chemical potential increases.
For a fixed  $V<V_c(\mu)$, the range of this phase  for increasing $\mu/2t$ depends on the strength of the attractive interaction $g$.

On the other hand for $\left(\left|\mu\right|/2t\right)<1$,
but for $V>V_{c}(\mu)$, there is a gapped non-superconducting phase, that corresponds
to a topological insulator ($TI$), as will be discussed below. This phase
is characterized by zero energy modes localized at the ends of the
chain for $\mu=0$. There are also localized modes
if $\mu \neq 0$ that have finite subgap energy.

Notice that the conditions for the existence of a gap are given by
Eq. (\ref{eq:condicao fechamento gap}). For instance, in the case
of strong hybridization and weak or no superconductivity, such that $1+\left(\Delta/2t\right)^{2}<(V/2t)^{2}$,
the system becomes gapless whenever $(\mu/2t) \geq 1$ and $V  > V_c(\mu)$.
This corresponds
to the phase $M$ in the phase diagram of figure \ref{fig1}b which is a normal
metallic or insulating phase (not shown in the figure) depending on the occupation number.

\begin{figure*}
\begin{centering}
\hfill{}%
\begin{minipage}[t]{0.33\linewidth}%
\includegraphics[scale=0.26]{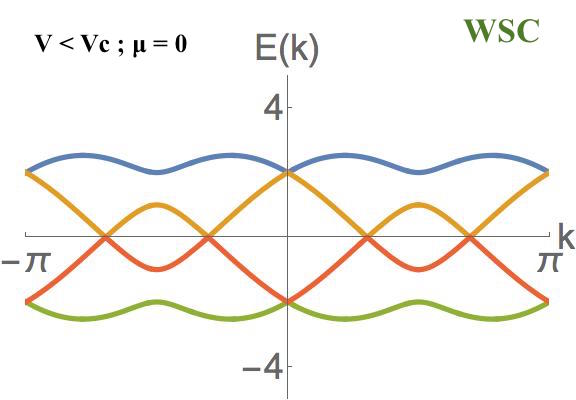}%
\end{minipage}\hfill{}%
\begin{minipage}[t]{0.33\linewidth}%
\includegraphics[scale=0.26]{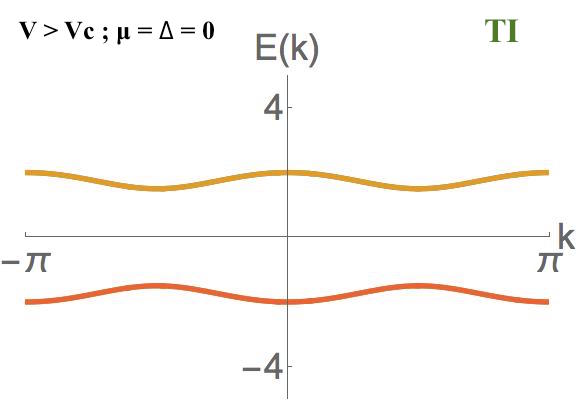}%
\end{minipage}\hfill{}
\par\end{centering}

\begin{centering}
\hfill{}%
\begin{minipage}[t]{0.33\linewidth}%
\includegraphics[scale=0.26]{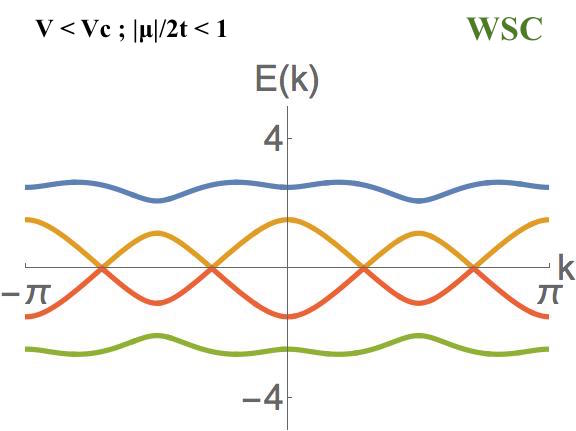}%
\end{minipage}\hfill{}%
\begin{minipage}[t]{0.33\linewidth}%
\includegraphics[scale=0.26]{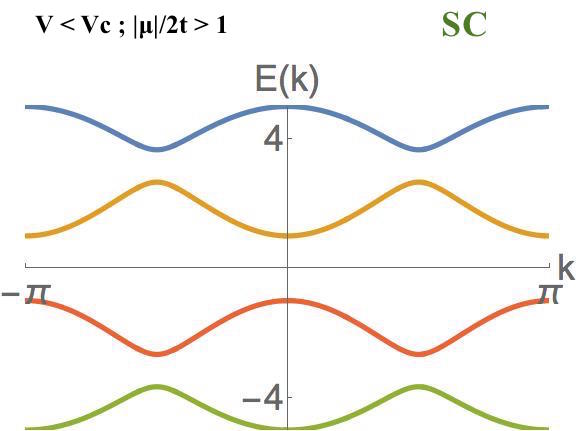}%
\end{minipage}\hfill{}
\par\end{centering}

\protect\caption{Energy spectrum in momentum space for different points on the phase
diagram shown in figure \ref{fig1}. The up left panel shows a system
with hybridization and superconductivity when $V<\Delta$, and the
chemical potential is fine tuned to zero (specifically, $t=1$, $\mu=0$,
$V=0.5$ and $\Delta=1.5$); the up right panel illustrate the case
where there is no self-consistent superconductivity (specifically,
$t=1$, $\mu=\Delta=0$ and $V=1.5$). The down panels show the cases
where there is superconductivity but the chemical potential is driven
out from zero. The left plot has the same conditions that the first
case but with $\mu=0.5$, and it illustrates the case with gapless
superconductivity. On the other hand, the right panel was set to same
conditions as before but $\mu=3$ and $\Delta=1$, and it illustrates
the full gapped superconducting region.}

\label{fig:energy spectrum momentum space}
\end{figure*}

In order to clarify the understanding of the phases discussed above,
in figure \ref{fig:energy spectrum momentum space} we plot the energy
spectrum in different regions of the phase diagram of figure \ref{fig1}b.
These figures illustrate the cases of appearance of gapped or gapless
superconductivity.

\subsection{Nature of the transitions}

\subsubsection{WSC-TI Half-filling }

Let us first consider the case of half-filling bands ($n=1$), where
the chemical potential is fixed at $\mu=0$. The constraint in equation
(\ref{eq:constrain}) now reads $\Delta_{k}^{2}>\left(\varepsilon_{k}^{2}+V_{k}^{2}\right)$.
Important points in momentum space correspond to those wavevectors
where this inequality becomes an equality, i.e.,
\begin{equation}
\tan^{2}\left(k_{0}\right)=\frac{\left(2t\right)^{2}}{\Delta^{2}-V^{2}}.
\end{equation}
At these points the system becomes gapless and they characterize the
Weyl points. From this result, we can immediately see that there
are no gapless nodes when $V>\Delta$. It is easy to see also that with
increasing hybridization the Weyl points collapse at $k_{0}=\pi/2 \equiv k_F$
for $V=V_{c}=\Delta$ at a discontinuous quantum phase transition
from the WSC to the TI phase where the superconducting order parameter
drops to zero. This collapse of superconductivity is associated with
the appearance of zero energy modes exactly at the Fermi $\mathit{surface}$
$k_{F}$, of the half-filled system. In the superconducting
side before the transition, the order parameter (when $\mu=0$) is given
by the gap equation,
\begin{equation}
\frac{1}{g}=\frac{4}{\left|\Delta\right|L}\sum_{k=0}^{k_{0}}\sin\left(k\right).
\end{equation}
where $k_{0}$ is the largest momentum value that contributes to the
superconductivity. Notice that this equation has no trivial analytic
solution since $k_{0}$ depends on $\Delta$. At the transition, for
$V_{c}=\Delta$ the momenta $k_{0}=\pm\pi/2$ and $\Delta_0=4g/2\pi$
before dropping abruptly to zero at the TI phase.

\subsubsection{WSC-SC}

We now investigate the transition from the non-trivial topological
superconductor to the trivial one by increasing the chemical potential
at fixed hybridization. Let us for simplicity consider the case of
$V=0$. The WSC-SC transition occurs for $(\mu/2t)_{c}=1$ as shown in figure~\ref{fig1}c.
It is associated, as can be easily checked with the collapse of two Weyl points at the center
of the Brillouin zone ($k=0$) and at its extremities ($k=\pm\pi$).
Expanding the dispersion relation of the excitations close to $k=0$
and $(\mu/2t)_{c}=1$, we get,
\begin{equation}
\omega_{2}(k)=2t\sqrt{(1-\frac{\mu}{2t})^{2}+(\frac{\Delta}{2t})^{2}k^{4}}.
\end{equation}
We have omitted the $k^2$ term, since its coefficient is proportional to $(1-\mu/2t)$ and vanishes at
the quantum topological phase transition at $(\mu/2t)_c=1$. Then, at the quantum critical point, the spectrum of excitations
$\omega_{2}(k)$$\propto k^{2}$, which allows to identify the dynamical exponent
$z=2$ for this transition. On the other hand at $k=0$, the gap $\omega_{2}(k=0)=(\mu/2t)_{c}-(\mu/2t)$,
vanishes linearly at the quantum critical point with a gap exponent
$\nu z=1$. The critical exponents $\nu=1/2$ and $z=2$ show that
the quantum phase transition from the topological
 to the trivial superconducting phase in the inter-band model is in a different
universality class from that of the Kitaev model. In the latter at
the QCP, $(\mu/2t)_{c}=1$ , the dispersion is linear implying a dynamic
exponent $z=1$. Since the gap vanishes linearly also, we get for the
correlation length exponent the value $\nu=1$ (see Ref.~\cite{Continentino-Caldas-Nozadze-Trivedi-2014}). These different values
of the critical exponents imply distinct behaviour for the compressibility
of the two models at the topological
quantum phase transition inside the superconducting phase. The compressibility close to this transition
is given by, $\chi_{c}=\partial^{2}f/\partial\mu^{2}\propto|((\mu/2t)_{c}-(\mu/2t)|^{-\alpha}$
where $f$ is the free energy density. The exponent $\alpha$ is related to the correlation length
and dynamical exponents by the quantum hyperscaling relation~\cite{Continentino-Japiassu-Troper-1989,Continentino-1989,Continentino-2001}, $2-\alpha=\nu(d+z)$.
It can be easily verified that while for the intra-band Kitaev model
$\alpha=0$, which is generally associated with a logarithmic singularity~\cite{Nozadze-Trivedi-2015}, for the inter-band model $\alpha=1/2$ implying an even stronger
singularity for the compressibility at the topological transition. Indeed in our model the topological transition is in the universality of the Lifshitz transition~\cite{Volovik-2009}.
Notice that this is a purely topological quantum phase transition, since both phases are characterised by the same order parameter. In spite of this, they have singularities described by critical exponents which obey the quantum hyperscaling relation~\cite{Continentino-Japiassu-Troper-1989}. Although the usual Landau approach of expanding the free energy in terms of order parameters that become small close to a continuous phase transition is of no use here, the renormalisation group still provides an adequate description of this critical phenomenon~\cite{Continentino-2001}.

\subsection{Fermi velocity}

\begin{figure*}
\includegraphics[scale=0.24]{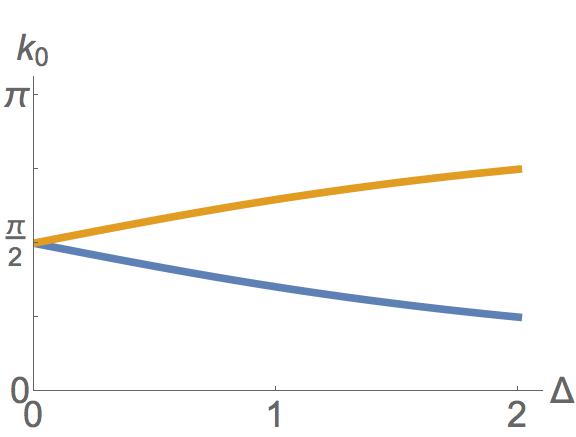}
\includegraphics[scale=0.24]{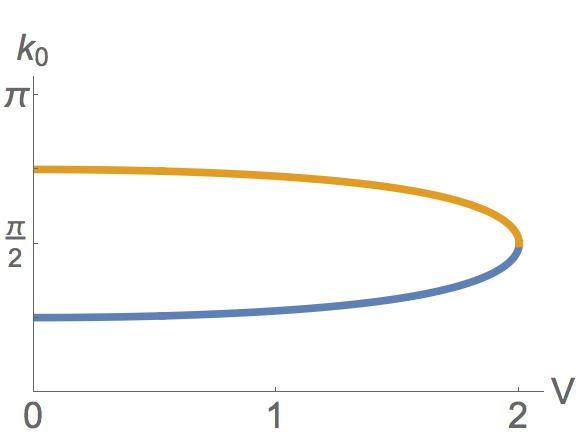}
\includegraphics[scale=0.24]{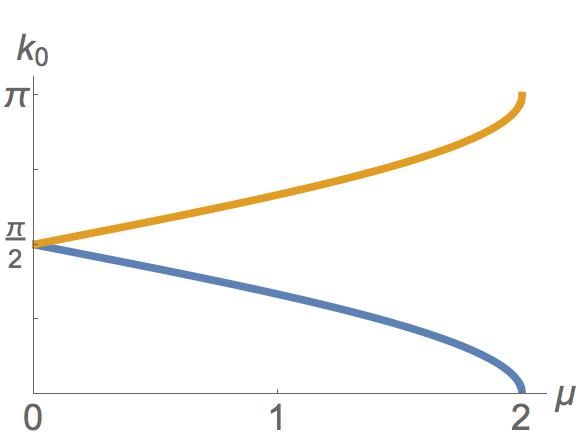}\\
\includegraphics[scale=0.24]{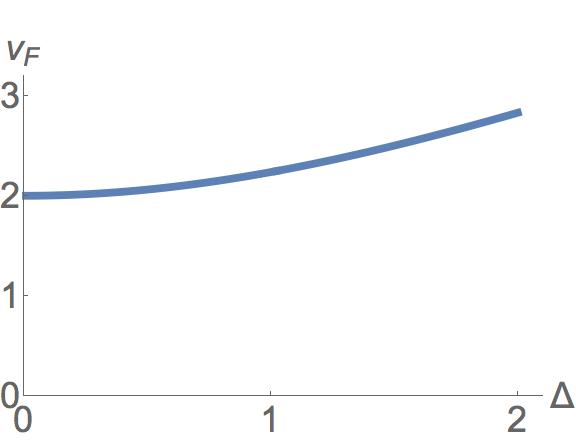}
\includegraphics[scale=0.24]{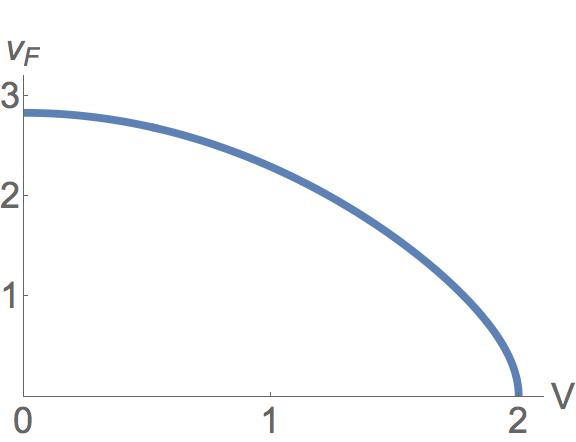}
\includegraphics[scale=0.24]{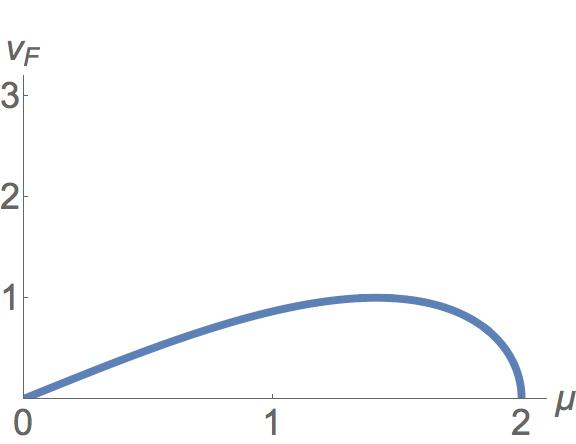}

\protect\caption{\label{fig2} Up row shows the positive Dirac points for a half-filled system ($\mu = 0$)
without hybridization ($V=0$), for a system with fixed superconductivity value
($\Delta=2$), and for fixed hybridization and superconductivity ($\Delta=V=2$) when increasing the chemical potential,
respectively from left to right. The down row shows the Fermi velocity
for each case, corresponding to its upper plot.}
\end{figure*}

We may calculate the Fermi velocity at the Fermi points $k_{0}$ expanding the energy spectrum in
equation (\ref{eq:dispersion relation}) in their vicinity.
The spectrum becomes $E\left(k\right)=E\left(k_{0}\right)+v_{F}\left(k-k_{0}\right)+O\left(k^{2}\right)$,
where $v_{F}=v_{F}\left(k_{0}\right)$ is the Fermi velocity. For
the general condition in equation (\ref{eq:constrain}) the Fermi
points are given by,
\begin{equation}
k_{0}=\arcsin\left[\sqrt{\frac{\left(2t\right)^{2}-\mu^{2}}{\Delta^{2}-V^{2}+\left(2t\right)^{2}}}\right],
\end{equation}
where, of course, the term inside the brackets must be within the
range $\left[0,1\right]$. The Fermi velocities
were obtained for three different situations and are shown in Fig.~\ref{fig2}. The up row  shows the variation of $k_0$ for a system
without hybridization ($V=0$), for a fixed $\Delta=2$, then for fixed $\Delta=V=2$, respectively from left to right.
The down row shows the Fermi velocities for each case, corresponding
to the upper plot.

\subsection{Energy spectrum in real space}

In order to find the energy spectrum in real space through the BdG
transformation, we write the Hamiltonian in the form,
\begin{equation}
\mathcal{H}=\bm{C}^{\dagger}\boldsymbol{H}\bm{C},
\end{equation}
where
\begin{equation}
\boldsymbol{C}=\begin{pmatrix}c_{1} & p_{1} & c_{1}^{\dagger} & p_{1}^{\dagger} & \cdots & c_{N} & p_{N} & c_{N}^{\dagger} & p_{N}^{\dagger}\end{pmatrix}^{T}
\end{equation}
and the matrix $\boldsymbol{H}$ is defined as
\begin{equation}
\boldsymbol{H}=\begin{pmatrix}\mathcal{H}_{11} & \cdots & \mathcal{H}_{1N}\\
\vdots & \ddots & \vdots\\
\mathcal{H}_{N1} & \cdots & \mathcal{H}_{NN}
\end{pmatrix},
\end{equation}
and is comprised by the following $\left(4\times4\right)$ interaction
matrices
\begin{equation}
\begin{cases}
\mathcal{H}_{r,r} & =-\mu\Gamma_{z0},\\
\mathcal{H}_{r,r+1} & =-t\Gamma_{zz}-i\Delta\Gamma_{yx}+iV\Gamma_{zy},\\
\mathcal{H}_{r,r-1} & =-t\Gamma_{zz}+i\Delta\Gamma_{yx}-iV\Gamma_{zy},\\
\mathcal{H}_{r,r'} & =0\qquad\forall\;r'\neq r,\;r+1\;\text{or}\;r-1.
\end{cases}
\end{equation}

The BdG transformation,
\begin{equation}
\begin{array}{c}
c_{r}=\sum_{n}\left[u_{s,n}(r)\gamma_{n}+v_{s,n}^{*}(r)\gamma_{n}^{\dagger}\right],\\
p_{r}=\sum_{n}\left[u_{p,n}(r)\gamma_{n}+v_{p,n}^{*}(r)\gamma_{n}^{\dagger}\right],
\end{array}
\end{equation}
diagonalizes the Hamiltonian, $\mathcal{H}=E_{0}+\sum_{n}E_{n}\gamma_{n}^{\dagger}\gamma_{n}$,
such that,
\begin{equation}
\boldsymbol{U}^{\dagger}\boldsymbol{H}\boldsymbol{U}=\boldsymbol{E},
\end{equation}
where $\boldsymbol{U}$ is formed by all the BdG coefficients $u_{s}$,
$v_{s}$, $u_{p}$ and $v_{p}$, and has the property to be unitary
$\boldsymbol{U}^{\dagger}\boldsymbol{U}=\mathbb{I}$. The matrix $\boldsymbol{E}$
is diagonal and contains the energy spectrum ($E_{n}$) of the system.

We have calculated the energy spectrum for a chain of $L=100$ sites
with two-orbitals per site and inter-band interactions in the presence
of hybridization. The spectrum consists of $4L$ energies. We have
checked that this size is large enough to prevent finite size effects.
We set the chemical potential to zero $\mu=0$, and take the hybridization
strong enough ($V>V_c$ or $\Delta =0$), such that the system is in the TI phase
of the phase diagram in Fig. \ref{fig1}b. We can see in Fig. \ref{fig:energy spectrum real space}a
the appearance of zero energy fermionic modes (see below). We also checked that these zero energy modes are localized at the ends of the chain, see Fig. \ref{fig:energy spectrum real space}b. This gapped insulating phase share the same properties of the topological insulating phase
found in a normal $sp$ chain \cite{Dzero-Sun-Galitski-Coleman-2010,Continentino-Caldas-Nozadze-Trivedi-2014,Alexandrov-Coleman-2014}.
In this situation the fermionic modes resemble the Majorana zero modes, as will be discussed in the next section.

Next, we remove the chemical potential from zero and keep ($V>V_c$),
but such that the system remains in the TI phase. We now find that
there are two (plus two particle-hole symmetric) energies in the spectrum
displaced from zero energy,
corresponding also to localized edge states but of finite energy and longer
spatial extent.

A more intriguing situation happens when we induce superconductivity on this TI phase
and other energies (for a small range of parameters)
displaced from the spectrum appear, see
figure \ref{fig:energy spectrum real space} down left panel.
The system now have 4 localized states (plus 4 particle-hole symmetric)
consisting of two double-degenerate states.
We also show that all those particular energies are localized in the end of
the chain, but with lower occupation number each, see down right panel of figure \ref{fig:energy spectrum real space}.

\begin{figure*}
\begin{centering}
\hfill{}%
\begin{minipage}[t]{0.33\linewidth}%
\includegraphics[scale=0.26]{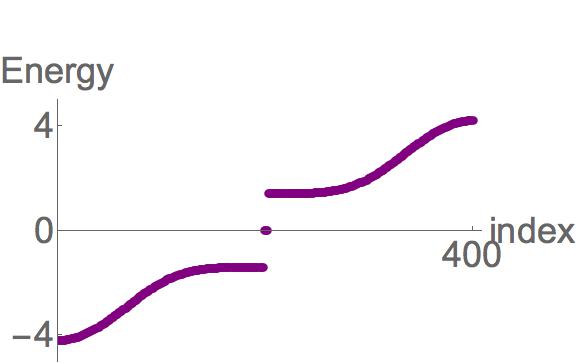}%
\end{minipage}\hfill{}%
\begin{minipage}[t]{0.33\linewidth}%
\includegraphics[scale=0.26]{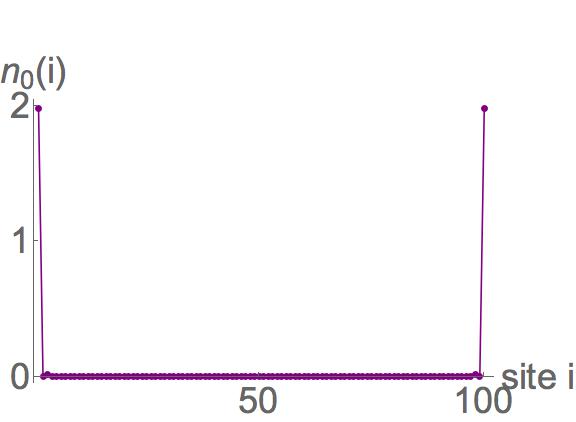}%
\end{minipage}\hfill{}
\par\end{centering}

\begin{centering}
\hfill{}%
\begin{minipage}[t]{0.33\linewidth}%
\includegraphics[scale=0.3]{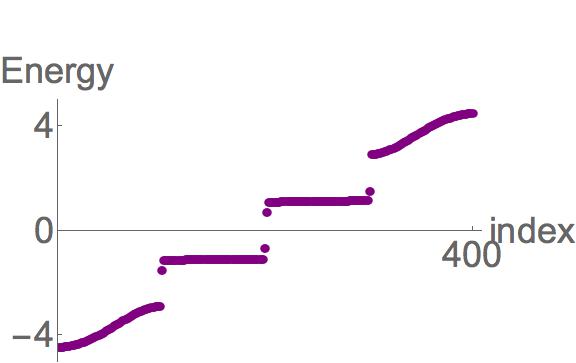}%
\end{minipage}\hfill{}%
\begin{minipage}[t]{0.33\linewidth}%
\includegraphics[scale=0.3]{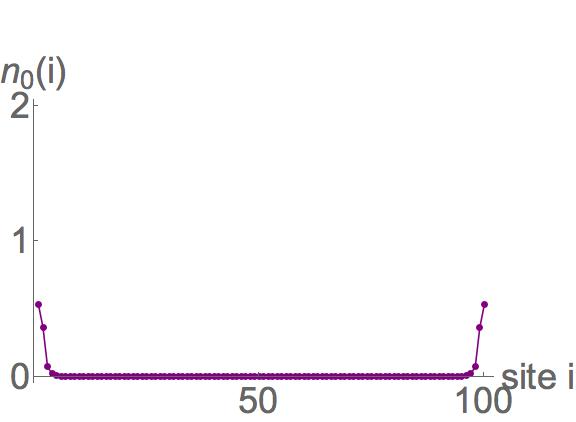}%
\end{minipage}
\hfill{}
\par\end{centering}

\protect\caption{Energy spectrum in real space for a chain of $L=100$ sites.
In the first row, the left panel shows a hybridised system with no, or weak induced, superconductivity
when the chemical potential is fine-tuned to zero, $\mu=0$.
The isolated point is four-fold degenerated.
On the right, the panel shows the localisation of the Majorana zero modes.
In the second row, the left panel shows the same system, with induced superconductivity,
but the chemical potential has its value near to zero.
Those isolated points are two-fold degenerated.
The right panel shows the occupation number on each lattice site for
one of those isolated energies. It is clear that they are localised
states at the edge.
}

\label{fig:energy spectrum real space}
\end{figure*}

\section{Topological invariants and edge modes\label{sec:Topological-condition}}

\subsection{WSC - topological invariant}

The winding number is a proper topological invariant that classifies the topological phase of a gapped 1D system.
In the gapless superconducting phase, it is not possible to calculate this by conventional methods since there are zero energy points that cannot be avoided in one dimensional systems,
or the sum over the Brillouin zone gives a vanishing winding number since the Fermi points appear in pairs and their
contributions cancel out \cite{Li-Chen-Drozdov-Yazdani-Bernevig-MacDonald-2014}.
On the other hand, let's look closer to one of the linear dispersion relations (see fig. \ref{fig:energy spectrum momentum space}) that crosses the zero energy at some point $k=k_0$.
In this region the Hamiltonian with Weyl nodes in 1D can be reduced to describe the two Bogoliubov bands that cross zero energy. The reduced low energy part of the Hamiltonian may be expanded in terms of Pauli matrix such that
\begin{eqnarray}
	H (k) = \sum_{i=1,2,3} d_i (k) \sigma_i.
\end{eqnarray}
Imposing inversion and charge conjugation operations \cite{Li-Chen-Drozdov-Yazdani-Bernevig-MacDonald-2014}
we end up with only one independent term, for example, $d_z(k) \sim (k-k_0)$,
where $\pm k_0$ are the two gapless points and the Hamiltonian is simply
\begin{eqnarray}
H & = & \left(k-k_{0}\right)\sigma_{z}
\nonumber \\
 & = & \delta_{k}\sigma_{z}.
\end{eqnarray}
For a positive chemical potential the energy at the Fermi level is positive,
which yields the eigenstate of matrix $\sigma_z$ to be $+1$ or $-1$ depending on the momentum $k$,
such that $\sigma_z \psi = -\psi$ if $k < k_0$ and $\sigma_z \psi = +\psi$ otherwise.
In this frame the winding number can be calculated as\footnote{Private discussions with B. Andrei Bernevig.}
\begin{equation}
	W=\text{sgn}\left(\delta_{k}\right)\left(\psi^{-1}\sigma_{z}\psi\right).
\end{equation}
Therefore, $W=+1$ indicates a non-trivial  phase  with topological excitations,  the Weyl nodes.
Note that if the system is not Weyl-like the Hamiltonian cannot be written in terms of one independent component and the winding number value is no longer preserved.
Furthermore, the above result is valid not only in the Pauli basis but also for higher order matrices,
provided that codimension is zero~\cite{Li-Chen-Drozdov-Yazdani-Bernevig-MacDonald-2014},
such as the $4\times 4$ Hamiltonian here considered.

Let us look for a special case of half-filling bands and no hybridization, $\mu = V = 0$.
In this situation we are in the WSC phase where the Weyl fermions appear at the momenta $k=\pm \pi /4$ and $k=\pm 3\pi /4$, when $\Delta \approx 2t$.
If we make a basis rotation on eq. (\ref{eq:hamilt k basis}) such that 
$\boldsymbol{C}_{k}^{\dagger}=\left(c_{k}^{\dagger}p_{-k}c_{-k}p_{k}^{\dagger}\right)$
it is easy to see that it can be decoupled in two $2\times 2$ Hamiltonians.
Near to the Weyl point $k_0$, one of these two Hamiltonians, e.g., the one for the basis $\left(c_{k}^{\dagger}p_{-k}\right)$, is
\begin{equation}
	H = \left( \alpha_0 (k - k_0) + m_0 \right)  \sigma_0 + \left( \alpha (k - k_0) + m \right) \sigma_y,
\end{equation}
with $\alpha_0 = m_0 = 2t/\sqrt(2)$ and $\alpha = m = \Delta/\sqrt(2)$.
Disregarding the mass term $m_0$ which leads to a shift on the energy,
we have only one independent term, the mass term $m$,
which cannot produce a gap by itself on the spectrum.

\subsection{TI - Majorana modes}
\label{TI - Majorana modes}

In order to clarify the existence of Majorana modes in our model, we write the Hamiltonian, Eq.~\ref{hamiltonian}, in real space. This is given by,
\begin{eqnarray}
\mathcal{H} & = & \sum_{i}\left\{ -\mu\left(c_{i}^{\dagger}c_{i}+p_{i}^{\dagger}p_{i}\right)+t\left(p_{i}^{\dagger}p_{i+1}-c_{i}^{\dagger}c_{i+1}\right)\right.\nonumber \\
 &  & +V\left(c_{i}^{\dagger}p_{i+1}-c_{i+1}^{\dagger}p_{i}\right)+\Delta\left(p_{i}c_{i+1}-p_{i+1}c_{i}\right)\nonumber \\
 &  & \left.+\text{h.c.}\right\} .\label{eq:Majorana real space}
\end{eqnarray}
This can be written in terms of Majorana operators, $\alpha_{A,r}$, $\alpha_{B,r}$,
$\beta_{A,r}$ and $\beta_{B,r}$, via the relations,
\begin{equation}
\begin{array}{c}
c_{r}=\frac{1}{2}\left(\alpha_{B,r}+i\alpha_{A,r}\right)\end{array}\quad\text{and}\quad\begin{array}{c}
p_{r}=\frac{1}{2}\left(\beta_{B,r}+i\beta_{A,r}\right)\end{array}.\label{eq:operadores-Majorana}
\end{equation}
Now, we perform a second transformation on Majorana fermions --
we call them unconventional hybridized Majorana fermions\footnote{We call them unconventional Majorana operators because they are not
valid for the same site, in other words, they do not obey the anti-commutation
relations when applied to the same site $r$, e.g., $\left\{ \gamma_{A,r}^{+},\gamma_{A,r}^{-}\right\} \neq0$
as it should be.} -- such that
\begin{eqnarray}
\alpha_{A,r}=\frac{\gamma_{A,r}^{+}+\gamma_{A,r}^{-}}{2\sqrt{\left(V-\Delta\right)}}, & \qquad & \alpha_{B,r}=\frac{\gamma_{B,r}^{+}+\gamma_{B,r}^{-}}{2\sqrt{\left(V+\Delta\right)}},\nonumber \\
\label{eq:Majoranas-nao-convencionais-2}\\
\beta_{A,r}=\frac{\gamma_{A,r}^{+}-\gamma_{A,r}^{-}}{2\sqrt{\left(V+\Delta\right)}}\quad & \text{and} & \quad\beta_{B,r}=\frac{\gamma_{B,r}^{+}-\gamma_{B,r}^{-}}{2\sqrt{\left(V-\Delta\right)}},\nonumber
\end{eqnarray}
the result is the following,

\begin{eqnarray}
\mathcal{H}' & = & \frac{i}{4}\sum_{r}\left\{ \left(1-\mathcal{C}_{t}\right)\left(\gamma_{B,r}^{-}\gamma_{A,r+1}^{+}-\gamma_{A,r}^{-}\gamma_{B,r+1}^{+}\right)\right.\nonumber \\
 &  & -\left(1+\mathcal{C}_{t}\right)\left(\gamma_{B,r}^{+}\gamma_{A,r+1}^{-}-\gamma_{A,r}^{+}\gamma_{B,r+1}^{-}\right)\nonumber \\
 &  & \left.-\mathcal{C}_{\mu}\left(\gamma_{B,r}^{+}\gamma_{A,r}^{+}+\gamma_{B,r}^{-}\gamma_{A,r}^{-}\right)\right\} ,\label{eq:Hamilt Majorana}
\end{eqnarray}
where $\mathcal{C}_{t}\equiv t/\sqrt{\left(V^{2}-\Delta^{2}\right)}$
and $\mathcal{C}_{\mu}\equiv\mu/\sqrt{\left(V^{2}-\Delta^{2}\right)}$.
If we go to the limit $\mu=0$ and take $\mathcal{C}_{t}=1$, such that the system is in the TI phase in Fig.~\ref{fig1}b ($V>\Delta)$, the Hamiltonian
reads,
\begin{equation}
\mathcal{H}'=-\frac{i}{2}\sum_{r=1}^{N-1}\left(\gamma_{B,r}^{+}\gamma_{A,r+1}^{-}-\gamma_{A,r}^{+}\gamma_{B,r+1}^{-}\right),
\end{equation}
which couples Majorana fermions only at adjacent lattice sites. Proceeding
with the same analysis as \cite{Alicea-2012}, we may easily see that
the Majorana modes $\gamma_{1}=\gamma_{A,1}^{-}$, $\gamma_{2}=\gamma_{B,1}^{-}$,
$\gamma_{3}=\gamma_{B,N}^{+}$, and $\gamma_{4}=\gamma_{A,N}^{+}$
are not present in the above Hamiltonian, it means that they have
no cost of energy to be added to the system; they are called Majorana zero
modes. In the present case we have two Majorana zero modes on each end,
and they can be combined to form one ordinary fermionic operator at each end as
\begin{equation}
f_{1}=\frac{1}{2}\left(\gamma_{1}+i\gamma_{2}\right)\quad\text{and}\quad f_{N}=\frac{1}{2}\left(\gamma_{3}+i\gamma_{4}\right),
\end{equation}
or, naively, we may think of two highly non-local fermionic operators such as
\begin{equation}
f=\frac{1}{2}\left(\gamma_{1}+i\gamma_{3}\right)\quad\text{and}\quad f'=\frac{1}{2}\left(\gamma_{2}+i\gamma_{4}\right).
\end{equation}
One may note that this result agrees with the conclusion on figure
\ref{fig:energy spectrum real space} up right panel. Moreover the Majorana zero modes
persists out of the fine-tuned $\mathcal{C}_{t}=1$, or $t^{2}=V^{2}-\Delta^{2}$,
provided
$\mu=0$, since we know from section \ref{sub:Energy-spectrum}
that there is no gap closing for this range of parameters.

When $\mu \neq 0$, the topological character is preserved, in the sense that we still have localized states at the ends of the chain,
but the Majorana zero modes are not robust such that they acquire a finite energy, see down left panel of figure \ref{fig:energy spectrum real space}.
In this situation, or when general referring, we call them by fermionic modes, instead of Majorana ones.

\subsection{TI - topological invariant}

The non-trivial topological character of the TI phase, can be shown by calculating the {\it winding number} for the special case $\mu=0$.
In this region of the phase diagram we have $\Delta=0$ and the $4\times 4$ Hamiltonian can be decoupled in two $2\times 2$ Hamiltonians, such as
\begin{equation}
\mathcal{H} = - V_k \sigma_y - \varepsilon_k \sigma_z.
\end{equation}
This equation can be rewritten as the Hamiltonian of a spin $1/2$ in a $k$-dependent magnetic field,
\begin{equation}
\mathcal{H} = - \mathbf{h}(k)\cdot \mathbf{\sigma},
\end{equation}
where $\mathbf{h}(k) = \left( h_x , h_y , h_z \right) = \left( 0, -V_k, -\varepsilon_k \right)$ with the properties $h_{x,y}(k)=-h_{x,y}(-k)$ and $h_{z}(k)=h_{z}(-k)$.
The  winding number $\nu$ is obtained as the product of the signs of the {\it magnetic field} on the center and at the extreme of the Brillouin zone, i.e.,
\begin{equation}
\nu = \text{sgn}(\mathbf{h}(k=0))\text{sgn}(\mathbf{h}(k=\pi)).
\end{equation}
Since $\mathbf{h}(k=0)=(0,0,2t)$ and $\mathbf{h}(k=\pi)=(0,0,-2t)$, we get $\nu=-1$, which characterizes the non-trivial topological character of the TI phase along the line $\mu=0$. The topological nature of this phase is associated with the existence of zero energy modes at the ends of the chain, as discussed above in sec. \ref{TI - Majorana modes} and also calculated and shown in figure \ref{fig:energy spectrum real space}.

As $\mu$ increases, we observe from the numerical solution that the end modes persist on the chain but acquire a finite energy. As $\mu/2t=1$, where the gap of the TI phase vanishes, they merge with the continuum of excitations.

If one calculate the winding number (by usual methods) for the whole system, the $4\times 4$ Hamiltonian including $\Delta$,
it shows itself trivial. The topology is hidden by the charge conjugation (or particle-hole) symmetry imposed to the system.
Some attempts to calculate the winding number using new methods were proposed to uncover this kind of topology.~\cite{Wakatsuki-Ezawa-Tanaka-Nagaosa-2014}.
On the other hand, the topological character of the whole TI phase is guaranteed
since it is adiabatically connected with the topological case just shown (when $\Delta = \mu = 0$).

\section{Conclusions\label{sec:Conclusions}}

We have studied in this work a 1D $sp$-chain with attractive inter-band  interactions and anti-symmetric hybridization due to the different parities of the $s$ and $p$ orbitals. The latter was shown to be responsible for the
appearance of topological phases~\cite{Continentino-Caldas-Nozadze-Trivedi-2014,Alexandrov-Coleman-2014} in non-interacting $sp$-chains.
We have shown that this model presents a rich phase diagram including non-trivial topological phases. It is interesting to compare it with the Kitaev model which also has an exact solution. In both models there is a weak coupling superconducting phase with non-trivial topological properties. However, while in Kitaev's model this phase is gapped, in our model it has Fermi points with gapless excitations.

We have studied the quantum topological phase transition between the weak coupling, non-trivial to the trivial, strong coupling superconductor and found that this transition in our model is in a different universality class from that of Kitaev's model.
In the strong coupling limit, the superconductivity disappears if $\mu$ is very large.
We have also shown the existence of a discontinuous quantum phase transition from a Weyl superconductor to a topological insulator with increasing hybridization. This is caused by the appearance of a zero energy mode exactly at the {\it Fermi surface} of the normal, non-interacting system.

We have shown that in the phase diagram of the present model there is a topological insulating phase, with zero energy fermionic modes at the ends of the chain. This phase has been characterized by calculating its winding number and the zero energy modes have been found both analytically and numerically.

The importance of models with exact solutions in the theory of topological matter has been now widely recognized. Besides throwing light in many exotic properties of these materials, they serve as guides for obtaining new types of excitations which are protected by topology. The present model, which as we have shown exhibits a rich variety of phases and different types of phase transitions, has many new features that allows for a deeper understanding of topological systems.

\begin{acknowledgments}
The authors would like to thank the CNPq and FAPERJ for financial
support and B.A. Bernevig and Griffith M.A.S. for useful discussions.
PDS acknowledges partial support from FCT through
project UID/CTM/04540/2013
\end{acknowledgments}

\bibliographystyle{apsrev4-1}
\bibliography{referencias}

\appendix

\section{Gap equation with hybridization \label{app:Gap-equation}}

In order to demonstrate the result in Eq. (\ref{eq:condition}) we
start calculating the gap equation
\begin{equation}
\Delta=-\frac{4g}{L}\sum_{k}i\sin\left(k\right)\left\langle p_{-k}c_{k}\right\rangle ,\label{eq:gap equation geral}
\end{equation}
that we solved using the fluctuation-dissipation theorem
\begin{equation}
\left\langle p_{-k}c_{k}\right\rangle =\frac{i}{2\pi}\int f\left(\omega\right)\left[\left\langle \left\langle c_{k},p_{-k}\right\rangle \right\rangle ^{r}-\left\langle \left\langle c_{k},p_{-k}\right\rangle \right\rangle ^{a}\right]d\omega,\label{eq:fluctuation-dissipation theorem}
\end{equation}
where $f\left(\omega\right)$ is the Fermi distribution. The retarded
and advanced Green functions are obtained from the Greenian $\left\langle \left\langle \boldsymbol{G}\right\rangle \right\rangle _{k}=\left(\mathcal{H}_{\omega}\right)^{-1}$,
with $\mathcal{H}_{\omega}\equiv\omega\mathbb{I}_{4\times4}-\mathcal{H}_{k}$,
such that

\begin{equation}
\left\langle \left\langle \boldsymbol{G}\right\rangle \right\rangle _{k}=\frac{\boldsymbol{F}_{k}\left(\omega\right)}{\det\left(\mathcal{H}_{\omega}\right)}.
\end{equation}
With the same basis used in equation (\ref{eq:hamilt k basis}) the
Green function $\left\langle \left\langle c_{k},p_{-k}\right\rangle \right\rangle $
is the $\left(4,1\right)$ element of the Greenian matrix, which can
be written as
\begin{eqnarray}
\left(\boldsymbol{F}_{k}\left(\omega\right)\right)_{4,1} & \equiv & -i\Delta_{k}F_{k}\left(\omega\right)\nonumber \\
 & = & -i\Delta_{k}\left[\left(\omega-\epsilon_{k}\right)^{2}-\mu^{2}-\Delta_{k}^{2}+V_{k}^{2}\right].
\end{eqnarray}
If one put this into the Eq. (\ref{eq:fluctuation-dissipation theorem}),
proceed with the integral calculation and takes the zero temperature
limit, after some calculations, will find that mean value of the operators
is
\begin{equation}
\left\langle p_{-k}c_{k}\right\rangle =\frac{-i\Delta_{k}}{2\left(\omega_{1}^{2}-\omega_{2}^{2}\right)}\left\{ \frac{F_{k}\left(-\omega_{2}\right)}{\omega_{2}}-\frac{F_{k}\left(-\omega_{1}\right)}{\omega_{1}}\right\} ,
\end{equation}
where
\begin{equation}
\omega_{1}(k)=\sqrt{Z_{1}+2\sqrt{Z_{2}}}\quad\text{and}\quad\omega_{2}(k)=\sqrt{Z_{1}-2\sqrt{Z_{2}}},\label{eq:geral-energias}
\end{equation}
with
\begin{eqnarray}
Z_{1} & = & \epsilon_{k}^{2}+V_{k}^{2}+\Delta_{k}^{2}+\mu^{2}\\
 & \text{and}\nonumber \\
Z_{2} & = & \left(\Delta_{k}^{2}+\mu^{2}\right)\left(V_{k}^{2}+\epsilon_{k}^{2}\right).
\end{eqnarray}
If we put this result into the gap equation we find that
\begin{eqnarray}
\frac{1}{4g} & = & -\frac{1}{L}\sum_{k}\frac{\sin^{2}\left(k\right)}{2\left(\omega_{1}^{2}-\omega_{2}^{2}\right)}\left\{ \frac{F_{k}\left(-\omega_{2}\right)}{\omega_{2}}-\frac{F_{k}\left(-\omega_{1}\right)}{\omega_{1}}\right\}.
\nonumber \\
\end{eqnarray}
We can rewrite the right hand side of the equation above as
\begin{equation}
\begin{split}\frac{1}{\left(\omega_{1}^{2}-\omega_{2}^{2}\right)}\left\{ \frac{F_{k}\left(-\omega_{2}\right)}{\omega_{2}}-\frac{F_{k}\left(-\omega_{1}\right)}{\omega_{1}}\right\} =\\
=\frac{1}{\left(\omega_{1}+\omega_{2}\right)}\left\{ \text{sgn}\left[A\left(k\right)-B\left(k\right)\right]-1\right\} ,
\end{split}
\end{equation}
where
\begin{equation}
A\left(k\right)=\left(V_{k}^{2}+\epsilon_{k}^{2}\right)\qquad\text{and}\qquad B\left(k\right)=\left(\Delta_{k}^{2}+\mu^{2}\right).
\end{equation}
We may write this result in the compact form
\begin{equation}
\begin{split}\frac{1}{\left(\omega_{1}^{2}-\omega_{2}^{2}\right)}\left\{ \frac{F_{k}\left(-\omega_{2}\right)}{\omega_{2}}-\frac{F_{k}\left(-\omega_{1}\right)}{\omega_{1}}\right\} =\frac{-2\delta_{k}}{\left(\omega_{1}+\omega_{2}\right)}\end{split}
,
\end{equation}
where\foreignlanguage{brazil}{
\begin{equation}
\delta_{k}=\begin{cases}
1 & \qquad\text{if}\qquad B\left(k\right)>A\left(k\right),\\
0 & \qquad\text{otherwise}.
\end{cases}
\end{equation}
}Such condition shows that the gap equation to this model is
\begin{equation}
\frac{1}{g}=\frac{1}{L}\sum_{k}\frac{4\sin^{2}\left(k\right)}{\left(\omega_{1}+\omega_{2}\right)}\delta_{k}.
\end{equation}

\end{document}